# Trends in deep learning for medical hyperspectral image analysis

**Uzair Khan [1], Paheding Sidike [2], Colin Elkin [1], and Vijay Devabhaktuni [1]**

[1]  Department of Electrical and Computer Engineering, Purdue University Northwest
[2]  Department of Applied Computing, Michigan Technological University

**Abstract:** Deep learning algorithms have seen acute growth of interest in their applications throughout several fields of interest in the last decade, with medical hyperspectral imaging being a particularly promising domain. So far, to best of our knowledge, there is no review paper that discusses the implementation of deep learning for medical hyperspectral imaging, which is what this review paper aims to accomplish by examining publications that currently utilize deep learning to perform effective analysis of medical hyperspectral imagery. This paper discusses deep learning concepts that are relevant and applicable to medical hyperspectral imaging analysis, several of which have been implemented since the boom in deep learning. This will comprise of reviewing the use of deep learning for classification, segmentation, and detection in order to investigate the analysis of medical hyperspectral imaging. Lastly, we discuss the current and future challenges pertinent to this discipline and the possible efforts to overcome such trials.

**Keywords:** Deep learning, neural networks, machine learning, medical image analysis, medical hyperspectral imaging, COVID-19

## 1. Introduction

Medical imaging refers to images used to aid in clinical work relative to the human body such as surgical procedures, diagnoses for impeding diseases, or simply to analyze and study body functions and is primarily based on radiological research. For the last couple of decades in particular, modern imaging techniques such as X-rays, magnetic resonance imaging (MRI), and ultrasound have had significant impacts on not only medical symptoms analysis but also on the spawning of more imaging techniques for improvised examination. Computed tomography (CT) scan, for example, is an X-ray procedure that displays the cross-sectional image of the body, which now also helps to assess brain or head-related injuries [1]. Another circumstance in which X-rays have been influential for further progress in research and application is Mammography, where a low energy x-ray photon beam is implemented to diagnose breast cancer, which is presently a common medical problem worldwide [2]. These implementations of x-rays are more commonly known as computer-aided diagnosis (CAD) [3]. More than often for medical imaging, image fusion is utilized, in which the final output image consists of more meaningful information, which is obtained from multiple input images [4]. An emerging imaging method in the discipline of biomedicine is Terahertz (THz) imaging, which is currently working to overcome its limitations with the aid of nanotechnology by finding its roots in the highly promising medical imaging methodology and becoming safer than traditional methodologies [5]. The primary goals for any effort to obtain medical images for diagnosis is via a non-invasive and inexpensive methodology. While the aforementioned techniques provide the desired output, each method is either invasive in some form, or not economical and in some cases, neither. This leads to professionals in the medical field and in academia alike to look towards alternative methods for imaging that are better suited for the required criteria of a non-invasive and a low-cost method.

Hyperspectral imaging (HSI) is a developing imaging technique amid the medical imaging modality and offers noninvasive disease diagnosis. HSI comprises various images aligned in adjacent narrow wavelengths or spectral bands (often in the range of hundreds) and recreates a reflectance spectrum of all the pixels in the band [6]. This is done by separating light using a spectral separator consisting of bandpass filter(s) and accumulating it on a focal plane detector (typically a complementary metal oxide semiconductor (CMOS) sensor) to form the image. Hyperspectral imaging has traditionally been used for remote sensing [7] [8], agriculture [9], food safety & quality assessment [10] [11], image enhancement [12], disaster monitoring [13] [14], feature extraction [15], classification [16], object detection [17] [18] [19], and recently even for conservation of art [20]. For medical imaging, it is principally obtained by targeting tissue samples by transmission of light and used to diagnose and detect various types of cancers [21] and other medical applications. With cancer being the second leading cause of death in the US, this significantly impacts the medical society and its research to eliminate cancer. Medical Hyperspectral Imaging (MHSI) has previously aided in successfully distinguishing between tumors and normal tissues from a rat breast tumor model, providing a clear indication of how influential MHSI can be for future research on breast cancer [22]. It has also proven its significance in detection of cancerous tissue cell detection from normal tissue specimen for neck and head cancer [23]. Gastric cancer is the most common cause of cancer in the United States, and its future diagnosis will also be significantly simplified thanks to MHSI [24]. Furthermore, it has also facilitated cancer detection of head and neck in surgical specimens [25]. All these MHSI applications for several medical diagnosis have led to the development of multiple algorithms to more accurately and efficiently classify the cancerous tissues from a sample [26], which further instigates on the different techniques for the processing and analysis of MHSI.

Artificial intelligence, and its most common subset known as Machine Learning, is a highly popular approach to process hyperspectral images and extract meaningful data from it. Machine learning (ML) algorithms make use of data and statistical models to learn and identify patterns to complete specific tasks and make decisions with or without human supervision. Several such ML algorithms are utilized when examination of hyperspectral images is considered and consequently in identifying and classifying differences in a tissue specimen when studying MHSI. While ML algorithms can be rudimentarily divided into supervised and unsupervised learning models, the algorithms are prone to develop into increasingly complex models as we delve further into deep learning (DL) [27], a branch of ML that is influenced by the structure and function of the human brain. In supervised learning, the model is fashioned from a dataset containing a number of input features and outputs, or labels. This model is formed by finding the optimal model parameters from a training sample of the dataset, which is subsequently used to predict the outcome based on the minimized cost function. Unsupervised learning processes data without any particular structure and is trained to find patterns and typically create groupings based on clusters.

Traditional ML algorithms typically used for MHSI applications lean towards classification models for identification and diagnosis, all of which include $k$-nearest neighbors (kNN) [28], linear discriminant analysis (LDA) [29], and support vector machines (SVM) [30] [31] [32] [33], with the latter being most prominently applied. With ML already heavily facilitating the processing of MHSI, the next logical step is to apply deep learning to achieve a more cost-effective and more accurate prediction model, which will provide a more comprehensive diagnosis to this pertaining medical issue. Deep learning in general has started taking off extraordinarily as early as 2012 [34] [35], and deep learning for MHSI has been no different in this regard. This subset of ML has already proven to outperform traditional ML techniques in the principle of head and neck surgery [36] and looks promising for MHSI in a broader aspect. Deep learning has also been applied for classification of a previously listed example related to head and neck cancer and shows significant improvement in accuracy over SVM and kNN [37].

Our end goal from this paper will be to highlight the key DL methods being used for MHSI and the challenges for successful application of DL in MHSI. This is because the DL methods to be implemented for MHSI in the coming decade will have a significant implication for future studies in several key areas of the medical discipline, including cancer research. The rest of the survey will be discussed as follows: Section 2 will introduce the fundamentals of DL methods that are currently being used as well as several emerging algorithms. Section 3 will discuss which DL methods are presently implemented for MHSI as well as additional algorithms that could be applied in the coming decade. Section 4 will ultimately discuss the current challenges faced by DL for MHSI.

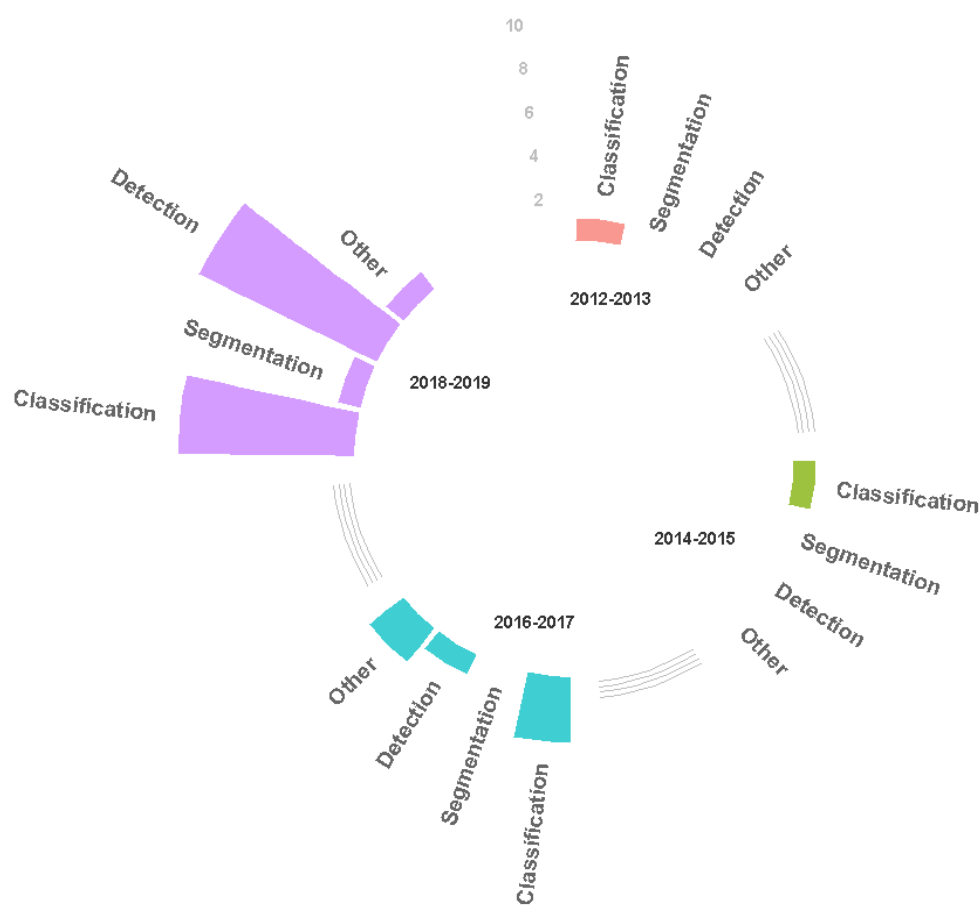

***Figure 1.*** *Chart representing the number of papers published in the last decade The bars depict the quantity of papers published spread apart evenly for every two consecutive years for each category of DL methodology.*

As previously mentioned, the trend in research for DL methods being used for MHSI has seen a sudden surge since as recently as 2012. This can easily be traced back to the ImageNet Large Scale Visual Recognition Challenge (ILSVRC) that was conducted in 2012, in which a deep learning method, which we will discuss later in Section 2, broke prior records by providing accuracy results close to 41% better than the best previous attempt [38]. Since then, more and more researchers have explored DL methodologies in a wide array of disciplines. Our survey covers papers published that specifically target the confrontation of medical hyperspectral imaging with deep learning techniques. For papers to be considered valid for our purpose, we examined keywords only with major publishers comprising of IEEE, Elsevier, Springer, SPIE, MDPI, International Journal of Biomedical Imaging, Journal of Biomedical Optics, and a handful of small publishers via Google Scholar as well as the publisher's respective search engines, with the latest being published in December 2019. While numerous publications are already present for deep learning being implemented for medical image analysis, with our topic of discussion being so highly specific, we cross-checked all sections within the publications that matched the subject matter of deep learning being implemented for medical hyperspectral imaging to verify papers for this survey. For certain situations in which published work corresponded with multiple papers, we only considered papers with greater significance in regard to their contributions. The goal for these papers being included in this survey, as mentioned earlier, encompassed the benefits of deep learning methodologies, how they are contributing towards medical hyperspectral imaging, and the challenges being faced for effectively applying these deep learning techniques to medical hyperspectral imaging.

With the rise of a global pandemic in the shape of coronavirus known as COVID-19 in late December of 2019, the race to find the solutions to its discovery and eventually its termination in form of a vaccine has severely heated up during early 2020. COVID-19 cases arise from cases of pneumonia, and as a result, affected humans display severe respiratory diseases and eventually death. In terms of applications of DL for medical imaging in regard to COVID-19, it is currently in the phase of being utilized heavily to detect the presence of COVID-19 amongst the screening of potential patients. The predominant testing technique currently being employed to detect COVID-19 is transcription-polymerase chain reaction (RT-PCR). X-ray procedures such CT scans are showing a pivotal role in early diagnosis, and it is in this domain that DL is providing significant results [39]. With several DL methodologies [40] [41] already cropping up to assist in the detection of COVID-19, it will not be long before DL also starts supplementing the eradication of the virus. For the purposes of this paper, however, we could not find any paper discussing DL in MHSI for the COVID-19 within our search criteria.

## 2. Deep Learning Methods

Machine Learning methodologies are typically divided into either supervised or unsupervised learning algorithms. A learning algorithm that makes use of a dataset comprising of a set of input features and an output label to obtain a predicting model is known as a supervised learning algorithm. The output label in the dataset for supervised learning could be categorized as either a classification or regression. A classification problem categorizes the output into discrete values such as type A and type B. A regression problem provides the output as continuous value, which could represent a real value such as dollars or a more symbolic numerical value, such as a normalization of a different number. The learning part in supervised learning refers to finding the optimal weights or parameters that minimize the cost (or loss) function. This would mean that the no further loss can be achieved and that the weights cannot be further improvised, which results in the best-fit model with the given inputs, bias, and learning rate. After successfully creating the model, a portion of the data is used to test the accuracy and/or the efficiency of the model. This is simply done by comparing the output from the model against the actual value from the dataset. While this is also implemented in the process of minimizing the cost function, the goal of testing the model is to perceive the vulnerability of the model.

An unsupervised learning algorithm is used to figure out the underlying structure of a dataset, which does not consist of a definite output. The algorithm does this by either clustering the output into categories or by finding the associative properties among the input parameters. Since the initial weights and parameters are selected ambiguously, the resulting output is not always similar every time, as a different model is obtained during every training process. In the following sections, we will delve upon various deep learning methods, which are built upon the ML fundamentals previously stated.

### 2.1 Supervised Learning Methods
### 2.1.1 Neural Networks

Neural Networks (also known as multilayer perceptron) are a category of learning algorithms built upon the idea and structure of a human brain, as the name suggests, and it lays the foundation for the majority of the deep learning methods. A neural network contains neurons (mathematically referred to as a perceptron) as the base unit which comprises of an activation number, and other sets of parameters such as weights W and bias b. This can be simplified to an activation function, which can be expressed as

$$a^{(1)} = \sigma(Wa^{(0)} + b), \tag{1}$$

where $a^{(0)}$ represents the initial activation numbers, or the input features, and $a^{(1)}$ refers to the activation numbers for the next layer. The $\sigma$ refers to the transfer function, which is traditionally denoted as either a sigmoid function or a step function. To form a fully constructed layer at each interval, we obtain the dot product between the weight vector and the input features vector. This represent a single layer of neurons, which has a simple feedforward mechanism, i.e. the neuron takes in a single input, performs the operation on it, and then passes it on to the next layer. A multi-layer perceptron (MLP) consists of two or more layers of neurons or perceptrons, which are also known as hidden layers. All of these layers from MLP combine together to form the basis of what is more commonly known as a Deep Neural Network (DNN). While there are different additional DNN techniques on which we will be expanding next, MLP is one of the most basic DNN architectures.

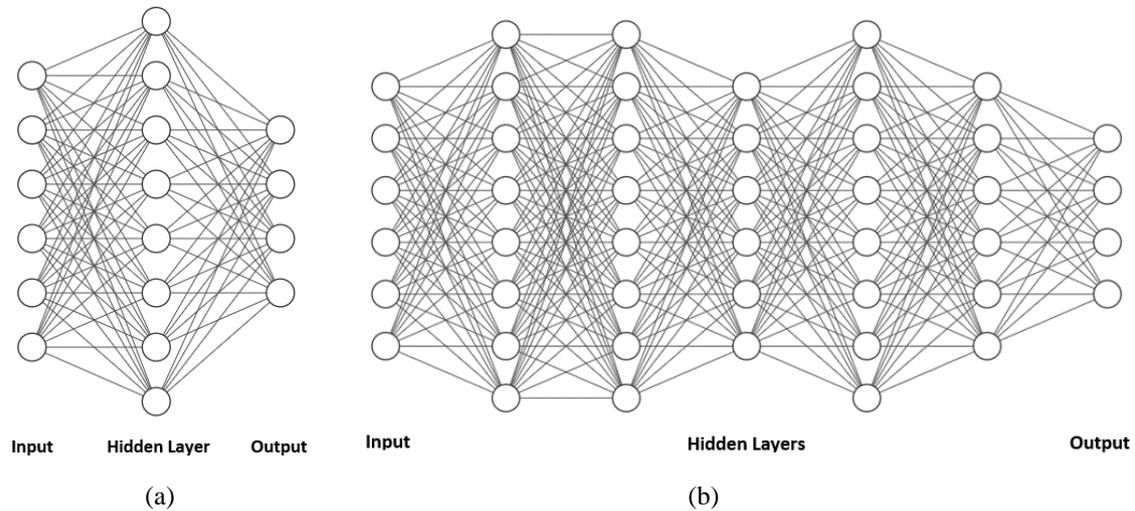

*Figure 2. Visual representation of (a) simple neural network: a machine learning algorithm modelled from the biological neural network of a human brain, (b) deep neural network: a neural network comprising of more than one hidden layer of neurons for its architecture.*

### 2.1.2 Convolutional Neural Networks

Convolution Neural Network (CNN) is an exceedingly popular deep learning algorithm that is used to classify an input image by recognizing patterns and features in order to differentiate between objects. In fact, one of the earliest perceptron concepts in 1957 was designed with the purpose of classifying the input image as man or woman [42]. The weights in a CNN architecture are designed to perform convolution operations on the input image rather than the features. In a convolution layer, the spatial structure is preserved along with any temporal dependencies with the usage of filters or kernels. This means that instead of appropriating a weight for an input feature via the combined stretched matrix from the input features, a small filter with appropriate dimensions relative to the input image is applied to perform the element-wise dot product between the filter and a chunk of the image otherwise 'convolved' in order to obtain the activation. Note that the depth of the filter should match the depth of the image to make this convolution possible. This process is repeated over all the spatial locations of the image, thereby providing us with a final activation map over the spatial region.

CNNs are predominantly part of typical classification architecture widely used in medical image analysis. One example for simpler architectures would be LeNet [43], which was introduced over two decades ago, which despite being a rather shallow network, displayed the basic concept of a CNN in a simple and elegant fashion, using the tangent function as the activation function. Later on, Alexnet [38] shattered expectations at the ILSVRC in 2012. Bearing similar characteristics to LeNet, AlexNet [38] also made use of kernels with a larger field of layers closer to the input and smaller kernels closer to the output, with the key difference being in the assimilation of rectified linear units (ReLU) for the activation function, which has since become the activation function of choice for modern CNNs due to higher classification accuracy. It can be duly noted that the rise in popularity of deep learning techniques also coincides after this series of events. This has led to the exploration of several architectures with farther reaching hidden layers. Building upon the base of deeper networks, more intricate layers are introduced which further reduces the error rate in a successful classification while also doing so more efficiently. In ILSVRC 2014, GoogleNet [44], used "inception" blocks that essentially reduced the number of operations performed at each layer by making use of smaller sets of convolutions. This inception module used the different sizes of convolution layers together, thereby allowing the final filter concatenation to stack the output together. Later, ResNet [45] was also introduced, which comprised of ResNet blocks. The residual block, as the name suggests, learns residual features, which in turn provide a shortcut that skips one or more layers. This further extended the efficiency of deeper models, thus enabling the more common vanishing gradient problem for deep learning models to be solved.

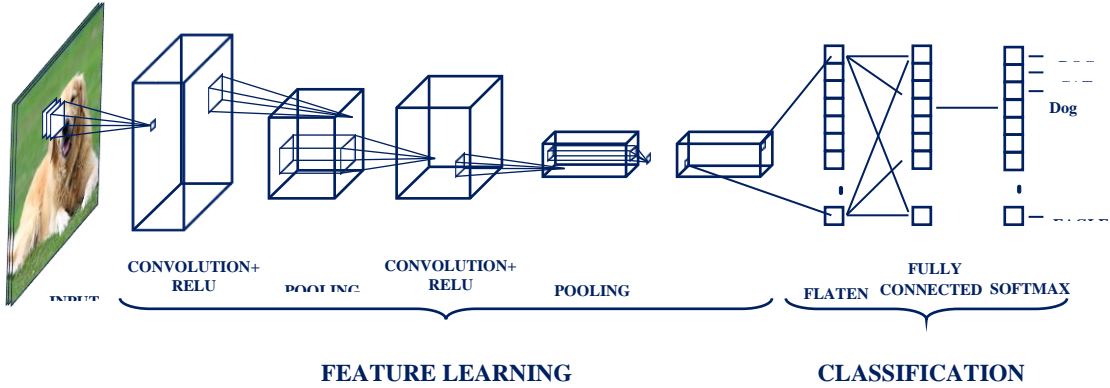

**Figure 3**. *Convolutional Neural Network (CNN): a deep learning architecture containing several convolutional and pooling layers, which are then connected at the output. This output in turn is used to classify the input image provided. CNN has proved exceeding popular in computer vision and visual image analysis as of late due to its high accuracy and precision.*

### 2.1.3 U-net

Segmentation is a popular architecture not only among the medical image analysis community but also for the field of computer vision in general. While CNNs classify the pixels of an image, segmentation provides inference by making prediction labels for each pixel that facilitates the enclosing of core object locations, thus separating the image into sections of fields. While this provides the unfortunate overlap of neighboring pixels over regions with the same convolutions being computed multiple times, several solutions have been proposed to overcome this ordeal, the most prominent of which is known as a Fully Convolution Network (FCN). The core idea behind FCN is to take the original CNN with arbitrarily sized input images and use the fully connected layer as convolutions to produce the segmented output. However, this still results in a degraded feature map due to the propagation through several pooling layers.

U-net architecture [46] provides a solution for this issue built upon the foundation of the FCN. This architecture consists of the basic FCN supported by an upsampling layer as opposed to a pooling layer, which concludes in an increase in resolution for the final output image. It uses stochastic gradient descent to train the network and calculate the energy equation with the aid of softmax, implemented pixel-wide across final feature map in which the softmax layer is defined as

$$p_k(x) = \frac{e^{a_k(x)}}{\sum_{k'=1}^{K} e^{a_k(x)}} \qquad , \tag{2}$$

where $a_k(x)$ is the activation function. This is then applied to an energy equation as

$$E = \sum w(x)\log(p_{k(x)}(x)), \tag{3}$$

where $w$ is the weight matrix for the model. This energy function is in short, a combination of the soft-max layer over the final feature map with the cross-entropy loss function.

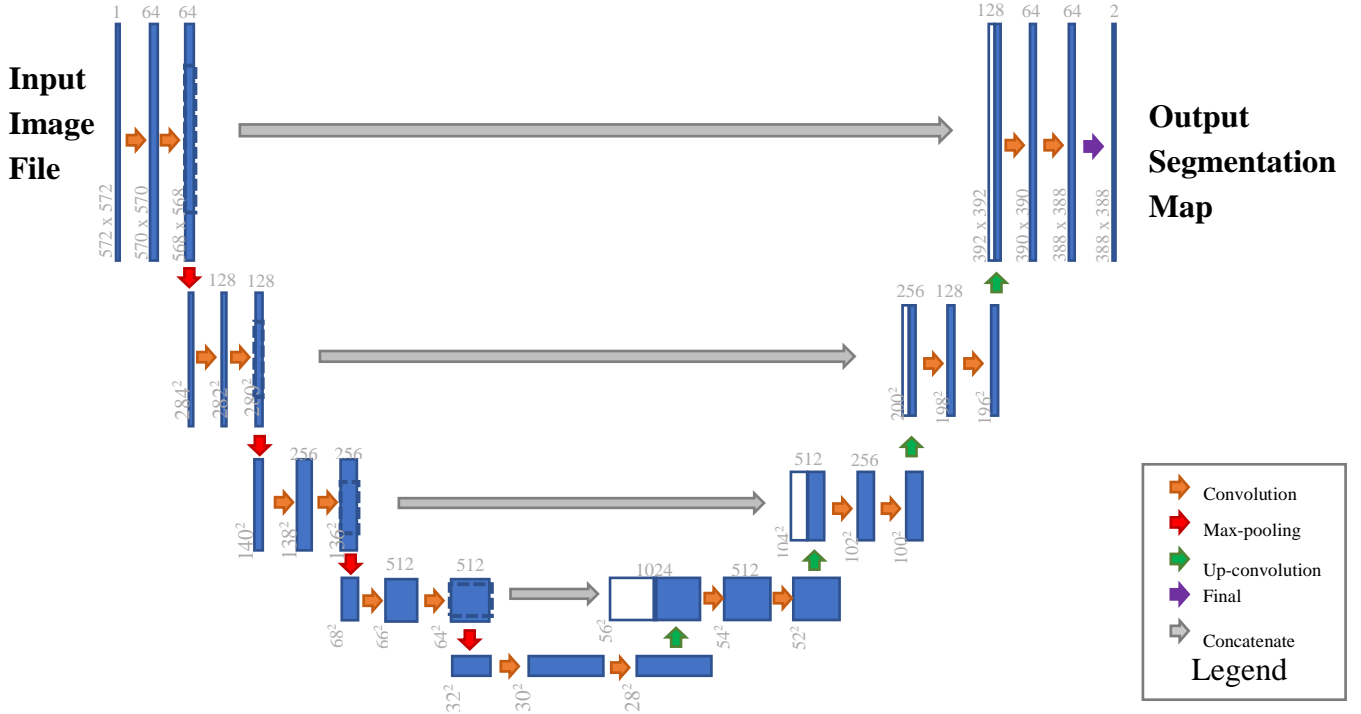

***Figure 3.*** *U-net architecture: a deep learning algorithm based upon Fully Convolutional Network (FCN) which in itself is built upon the foundation of CNN. Majority of the segmentation techniques involve u-net in one way or another, with it forming the core component of a proposed model for segmentation.*

### 2.1.4 Recurrent Neural Networks

Recurrent Neural Networks (RNNs) were developed with the thought of tackling progression of vectors over time, which is something CNNs cannot accomplish, as they are restricted to handling fixed vector size to provide fixed-size outputs. In addition, CNNs operate on a fixed number of layers. RNNs, by contrast, can have input and output vector of varying lengths, which make them invaluable for undertaking problems in the Natural Language Processing (NLP) domain [47], as the input matrix in such an application is constantly evolving. Another way in which the RNNs differ from traditional CNNs is that they are not feedforward systems and instead loop the outputs of each hidden layer back to itself. For a classification problem, the output from the hidden layer is used as the inputs along with the normal input for the hidden layer. This can be represented as

$$h_t = \sigma(W x_t + V h_{t-1} + b), \tag{4}$$

where $x_t$ is the input vector, $h_t$ is the hidden layer vector, W and V are the weight matrices, and $b$ is the bias vector.

The RNN, however, also experiences the same vanishing gradient problem during training as regular DNNs. Several solutions involving special memory units have been proposed for RNNs, with the Long Short Term Memory (LSTM) cell [34] being one of the most popular ones as well as having the distinction of being of the earliest.

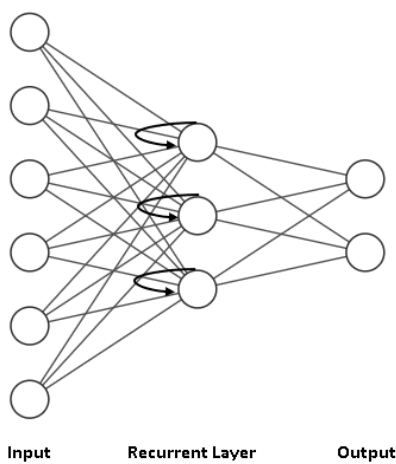

**Figure 5.** *A simple Recurrent Neural Network (RNN): a neural network in which the previous output is also used as inputs to modify the state of the hidden layer neuron. RNN has the capability of processing inputs of varying lengths without change in the model size, which is invaluable for constantly evolving datasets.*

## 2.2 Unsupervised Learning Methods

### 2.2.1 Auto-encoders

Autoencoder is a type of neural network that is primarily trained to provide a learned representation of the input. In other words, an autoencoder generates a replicate for the provided input after undergoing a handful of operations. These operations are performed over a single hidden layer in which the model 'encodes' and then 'decodes' the input to provide the mapped output. Although this process might seem meaningless on the surface, this gives us the opportunity to map how data is projected for a lower dimension. This is because the hidden layer has the smallest dimension in this network, and it also encompasses all the information to reconstruct the output for a same class of input. This is particularly useful in the case of anomaly detection, in which the autoencoder is fed the inputs from same class of data. Since the mapping features would produce an incoherent result with respect to the autoencoder hidden layer, the anomalies would be discovered easily.

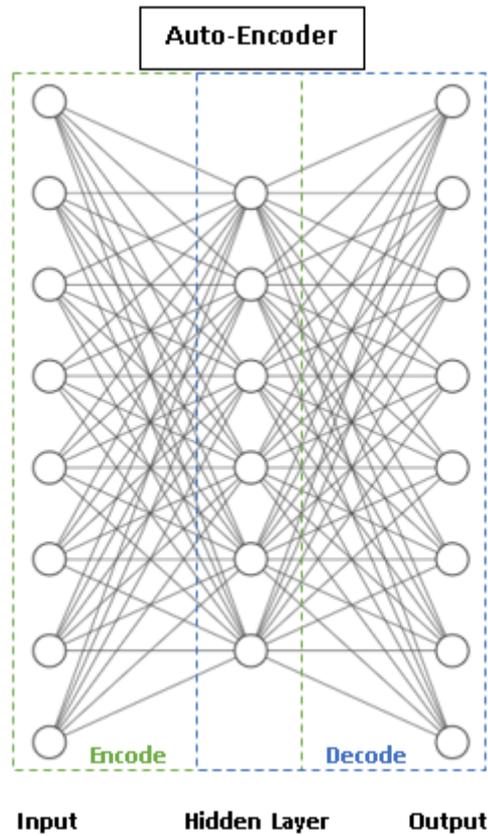

**Figure 4.** *Comprehensive layout of an autoencoder: a type of neural network that is used to understand the efficient data coding, which is particularly useful for dimension reduction for a given dataset.*

### 2.2.2. Restricted Boltzmann Machines

Restricted Boltzmann Machines (RBMs) are a relatively simpler deep learning system comprising of two layers: input and hidden. This forms the basis of deep belief networks. The nodes (neurons) from each layer form inter-layer connections while 'restricting' intra-layer connections, which helps derive its name. The RBM comprises of bidirectional communications between layers and is thus a generative model that uses the hidden layer to fashion new data points. It does so by defining an energy function for a state of input and visible units of (y,z) as

$$E(y,z) = \sum_i a_i y_i - \sum_j b_j z_j - \sum_i \sum_j y_i w_{i,j} z_j, \tag{5}$$

where $a$ and $b$ are the biases, $w$ is the weight matrix, and $y$ and $z$ are the states for hidden unit $j$ and visible unit $i$. The pair of possible hidden and visible vector is computed by finding the probability as

$$p(y,z) = \frac{1}{z} e^{-E(y,z)}, \tag{6}$$

where Z is known as a partition function. The RBMs are primarily used to pre-train a NN to generate the initial weights and then are used to form the foundation for other deep learning methods such as a Deep Belief Network (DBN). These DBNs are then used for many different applications, including cyber security [48], NLP [49], and of course medical image analysis [50].

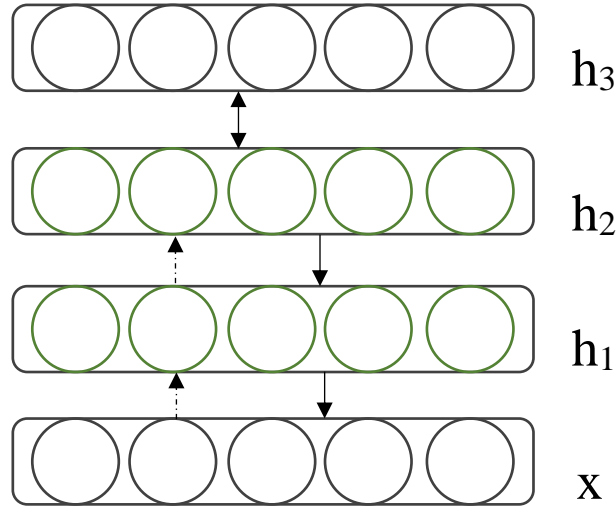

*Figure 5. RBM block diagram: A stochastic neural network that uses bidirectional layers (and consequently generative), comprising of hidden layers (h1, h2, & h3) and input layer x to effectively learn probability distributions for the given set of inputs.*

### 2.2.3. Generative Adversarial Networks (GANs)

GANs, used widely in image, video and audio generative scenarios [51], are a generative architecture (as the name suggests) based largely on probability-related setups for unlabeled datasets, which provide a better substitute to maximum likelihood estimators. A GAN architecture pits two neural networks against one another, with the purpose of generating synthetic labels that are comparable to actual data. In each iteration run, the two neural networks keep improving repeatedly at the required task. This procedure continues until the output from the generator resembles the actual sample data as closely as possible.

The generator network and the discriminator network competing against each other, as displayed in Fig. 8, can be considered the two supposed neural networks that highlight the concept of GAN as an example. The example can be considered as a min-max situation, in which the function V (D,G) can be described as

$$min_G max_D V(D,G) = E_{x \sim P_{data(x)}}[\log(D(x))] + E_{z \sim P_{data(z)}}[\log(1 - D(G(z)))]. \quad (7)$$

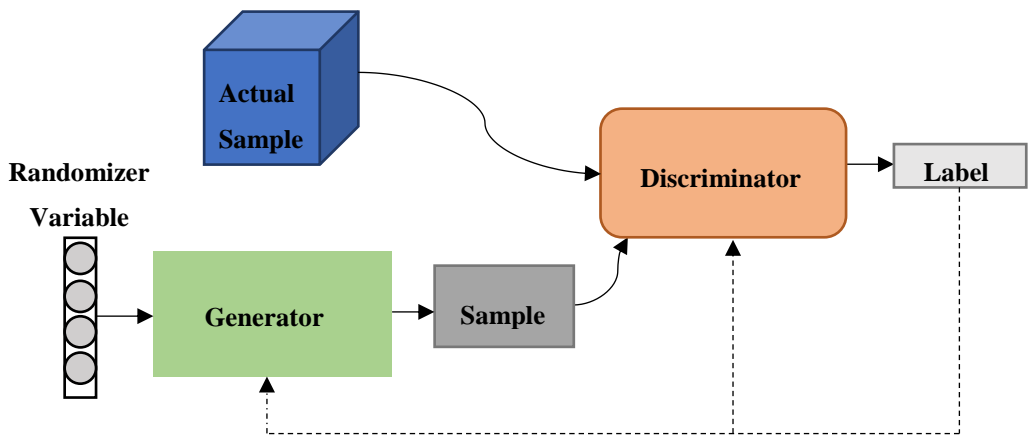

*Figure 6. Generative Adversarial Network block diagram: an architecture that pits two networks against each other, which is used to eventually learn the patterns of the input data in order to produce an output as closely as possible to an actual sample of the input.*

## 3. DL Methods for MHSI

### 3.1 Classification

Classification for pathological images has been one of the earliest fields of not just MHSI, but medical analysis in general, in which deep learning techniques have been a significant influence. Typical procedures for processing the MHSI using classification include applications such as cell classification [52] in order to identify and classify cancerous cells. This is also sometimes synonymous with detection techniques on the surface, in which the architecture is designed to detect traces of cancer from samples, as opposed to just classifying the cells themselves, which we will discuss in the next section. Classification techniques for MHSI often make use of transfer learning, which has proven to be quite useful in these scenarios. MHSI typically utilize small datasets comprising of medical diagnosis images (typically hundreds or thousands), as opposed to the field of computer vision in which the datasets could consist of millions of sample images, which is one of the reasons why transfer learning has been feasible for this discipline.

Transfer learning is fundamentally the use of a pre-trained network to classify input images from testing samples, and as mentioned earlier, the smaller datasets allow greater practicality and ease of use, as opposed to training the network to obtain new input features every time for virtually similar input image datasets. The implementation of deep learning for MHSI took some time to catch on, compared to deep learning being used for other areas of research. Earlier implementations of deep learning in MHSI began with classification of benign & malignant tissues or cell samples using ANNs, typically MLPs [21]. A particular study that explores HSI for characterizing kidney stones [53] made use of Principal Component Analysis (PCA) to determine appropriate variables to be used for a simple ANN model comprising of a hidden layer with four nodes, which was used to classify the type of kidney stone from the HIS, whereas a similar approach was also undertaken to classify different cancer types [54] [55]. There have been situations in which ANNs provided inferior results, which in turn highlights the problems related to deep learning, namely the absence of larger datasets. This particular study assessed the performance between four supervised algorithms that pitted ANN against random forest, SVM and k-nearest neighbors [56]. The paper found that from the 11 patient samples analyzed, SVM produced the best results, although they did remark that a larger training dataset may lead to better performance in general.

In more recent years, however, CNNs have become the prevalent choice for the task of tissue/cell classification. For instance, one paper used PCA for transfer learning for CNNs with kernel fusion [57] to complete the task of classification in MHSI. This publication made use of the Gabor kernel (which is implemented to obtain spatial features) and the CNN kernel to improvise conventional CNN execution for MHSI classification, with the proposed model showing improved performance. CNN has also been implemented to classify blood cell MHSI in a similar fashion [52] [58] [59], where increased pixel size for the MHSI produced better results with respect to classification accuracy, thus further proving the potential for CNN in MHSI. The CNN model showed promising prospects while classifying head and neck cancer [37], even for an animal model [60]. A proposed CNN model also produced successful results for classification of oral cancer diagnosis [61], the performance of which was verified by implementing the same dataset for SVM and DBN. The majority of the proposed CNN models discussed in this section all build upon the traditional architecture for CNN, which better suits the requirements for the medical diagnosis under examination.

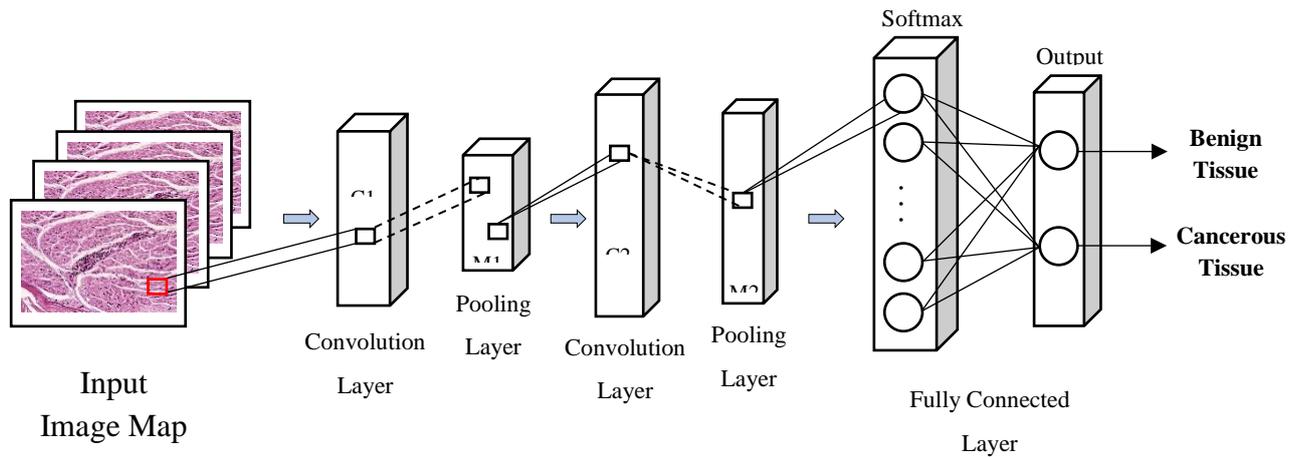

*Figure 7.* Visual representation of classification of a cancerous tissue sample.

### 3.2 Detection

Detection techniques in general refer to object detection for a given input image. For MHSI, detection techniques are habitually used for the purpose of detection of malignant cancerous samples, although they are also used for reconstruction of tissue samples. While ANN has been used to detect such cancerous samples [62], the majority of detection techniques implemented for MHSI use CNNs for classification of the pixels of the image, which is then used to detect the malign cancer cells/tissues [63] [64]. From this, after the pixel-wide classification is obtained from CNNs, it can also be considered as object classification, which is then used for post-processing to detect the presence of cancer or other purposes for the given input sample. One similar study implemented a CNN-based model to reconstruct tissue surface using an endoscopic probe, which displayed potential for practical applications [65]. One study also applied FCN to investigate tissue surface samples using an endoscopic probe in a similar fashion [66].

In recent years, CNN has been particularly useful for MHSI in head and neck squamous cell carcinoma detection [67] [68], although AEN has also been applied for a similar scenario [69]. Implementation of CNN was also observed for aiding brain tumor resection surgeries in real-time amongst the considered papers [70] and similarly towards determining skin cancer detection [71]. Such publications heavily suggest that although CNN is finely suited for classification techniques, as discussed in prior subsection, it also possesses potential for detection techniques that support a promising outlook for future research of detection techniques in MHSI.

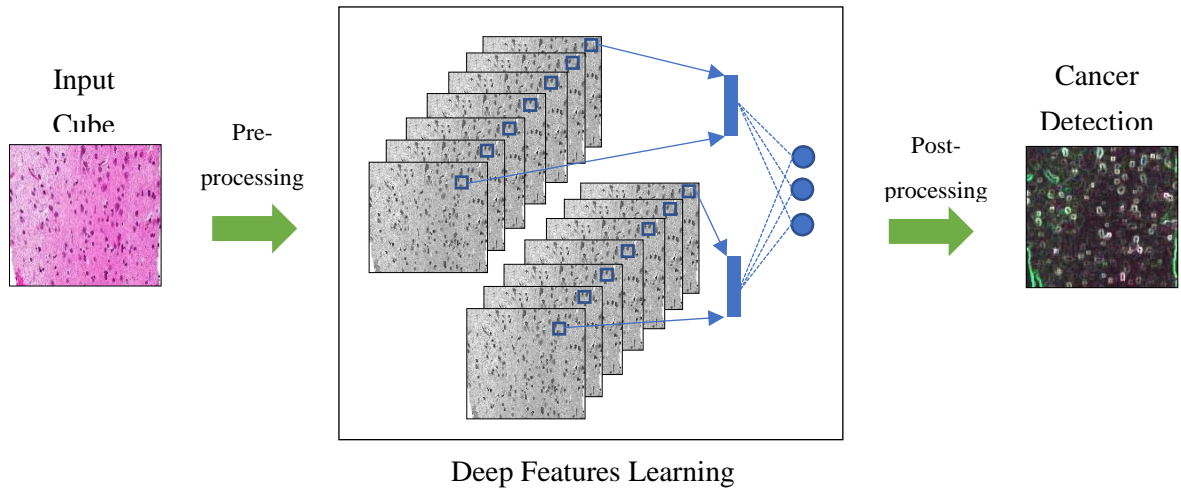

Input
Cube

Pre-
processing

Cancer
Detection

Post-
processing

Deep Features Learning

*Figure 8. Standard workflow for cancer detection using deep learning on medical image analysis.*

### 3.3 Segmentation

Segmentation provides outlines for certain parts of images that dictate the position, volume, and size of the relative objects in an image, as discussed in Section 2 of this study. For medical image diagnosis, this is particularly useful, as it could clearly outline certain organs, or noteworthy parts of a medical image. This is significant for medical diagnosis in which brain, liver or other important organs need to be distinguished from a medical image. While segmentation has been extensively used for medical diagnosis over the years [72] [73], in the case of MHSI, we could only find one notable paper that used a segmentation technique, specifically for the purpose of retinal image analysis [74].

This paper implemented a dense-FCN (FCN being the foundation for U-net [46]) to segment the retinal image. With the usage of *k*-means clustering on the input data, the study was able to lessen the complexity in order to aid the segmentation process and ultimately complete validation against alternate approaches involving other ML algorithms such as SVM and random forest. Its findings suggests that spectral data may provide the opportunity for improvised segmentation results for optic disc and macula segmentation, which is important for retinal imaging analysis, thereby prompting the urge for further research for segmentation techniques for MHSI.

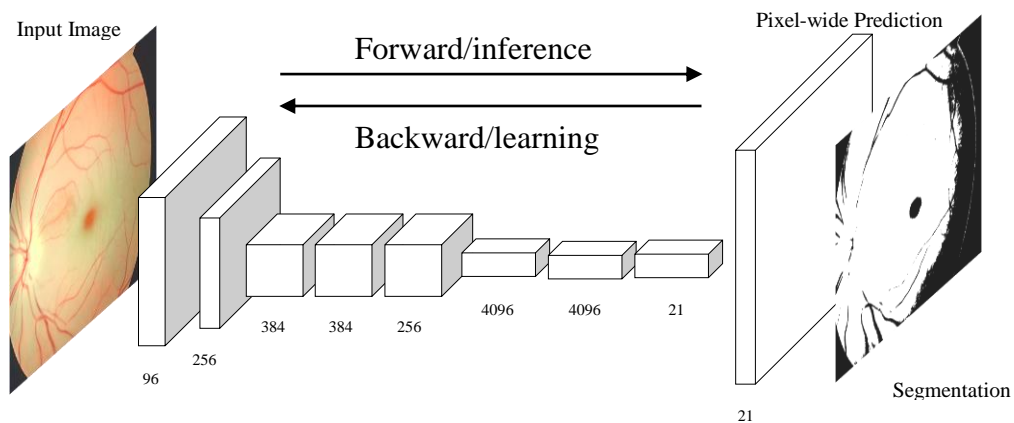

Input Image

Forward/inference

Backward/learning

Pixel-wide Prediction

96
256
384
384
256
4096
4096
21

Segmentation

*Figure 9. Classical representation of an FCN for the purposes of segmentation.*

3.4 Other

DL applications of MHSI go beyond the normal implementation of traditional domains of machine learning. This can be observed in a handful of papers considered in this study. For example, one study utilized GAN with the aid of an autoencoder using MATLAB ML tools in order to determine the tissue oxygen saturation using hyperspectral input data, the results of which are useful for the detection of ischaemia, a fatal disease that affects the blood supply to different organs of the body, particularly the heart muscles [75].

Another similar case in which blood oxygenation is determined using MATLAB tools was discussed in a 2017 study [76]. In this circumstance, the NN fitting tool was used to retrieve the parameters necessary to determine the oxygen saturation levels. Lastly, in another 2017 study, GAN was employed again to assist in the task of staining lung histology imaging, which was then used to further study the necessary tissue samples [77]. Overall, these studies suggest a broader field of applications of DL for MHSI in the near future. Tables 1 and 2 below consist of the relevant publications we obtained for this paper. Table 1 details the titles of the papers and the category of DL the paper pertains to, while Table 2 lists the application area of the papers.

Table 1. Papers using deep learning techniques for MHSI.

| Publication Title | Category |
|---|---|
| Cell classification using convolutional neural networks in medical hyperspectral imagery [52] | classification |
| Towards virtual H&E staining of hyperspectral lung histology images using conditional generative adversarial networks [77] | other |
| Hyperspectral image segmentation of retinal vasculature, optic disc and macula [74] | segmentation |
| Hyperspectral imaging for cancer detection and classification [54] | classification |
| Dual-modality endoscopic probe for tissue surface shape reconstruction and hyperspectral imaging enabled by deep neural networks [65] | detection |
| Probe-based rapid hybrid hyperspectral and tissue surface imaging aided by fully convolutional networks [66] | detection |
| Hyperspectral system for imaging of skin chromophores and blood oxygenation [76] | other |
| Medical hyperspectral imaging: a review [21] | classification |
| Deep convolutional neural networks for classifying head and neck cancer using hyperspectral imaging [37] | classification |
| Deep learning based classification for head and neck cancer detection with hyperspectral imaging in an animal model [60] | classification |
| Convolutional neural network for medical hyperspectral image classification with kernel fusion [57] | classification |
| Blood cell classification based on hyperspectral imaging with modulated Gabor and CNN [58] | classification |
| Medical hyperspectral image classification based on end-to-end fusion deep neural network [59] | classification |
| Tissue classification of oncologic esophageal resectates based on hyperspectral data **[56]** | classification |

Table 1 (continued)

| Publication Title | Category |
|---|---|
| A dual stream network for tumor detection in hyperspectral images **[62]** | detection |
| Adaptive deep learning for head and neck cancer detection using hyperspectral imaging **[69]** | detection |
| Hyperspectral imaging of head and neck squamous cell carcinoma for cancer margin detection in surgical specimens from 102 patients using deep learning **[67]** | detection |
| Computer-assisted medical image classification for early diagnosis of oral cancer employing deep learning algorithm **[61]** | classification |
| Hyperspectral imaging for head and neck cancer detection: specular glare and variance of the tumor margin in surgical specimens **[68]** | detection |
| Cancer detection using hyperspectral imaging and evaluation of the superficial tumor margin variance with depth **[64]** | detection |
| Surgical aid visualization system for glioblastoma tumor identification based on deep learning and in-vivo hyperspectral images of human patients **[70]** | detection |
| Optical biopsy of head and neck cancer using hyperspectral imaging and convolutional neural networks **[63]** | detection |
| Convolutional neural networks in skin cancer detection using spatial and spectral domain **[71]** | detection |
| Estimation of tissue oxygen saturation from RGB images and sparse hyperspectral signals based on conditional generative adversarial network **[75]** | other |
| Design of a multilayer neural network for the classification of skin ulcers' hyperspectral images: a proof of concept **[55]** | classification |
| Hyperspectral imaging based method for fast characterization of kidney stone types **[53]** | classification |

Table 2. Deep learning methodologies used in MHSI and their applications.

| Methodology | Application |
|---|---|
| CNN | Applied CNN with comparison to SVM for blood cell discrimination [52] |
| GAN | Virtual staining for lung histology images using GANs [77] |
| FCN | Segmentation for the automatic retinal image analysis [74] |
| ANN | Implementation of ANN and SVM for cancerous cell HSI [54] |
| CNN | CNN-based model used for endoscopic image reconstruction to enhance surgical guidance [65] |
| FCN | Spot detection for surgical endoscopic tissue sample aided by FCN [66] |
| NN | Matlab NN toolkit used to characterize of skin using 2d mapping of skin chromophore distribution [76] |

| ANN | This review paper discusses the use of ANN for classification for MHSI [21] |
|---|---|
| CNN | Classifying cancerous tissue samples from neck and head regions using CNN [37] |

Table 2 (continued)

| Methodology | Application |
|---|---|
| CNN | Detection of neck and head cancerous cells via classification using CNN [60] |
| | Improvisation of CNN using kernel fusion implemented for cell classification [57] |
| | Implementation of CNN for blood cell classification [58] |
| | Two-channel CNN for solving limited-samples problem for CNN models [59] |
| MLP | Evaluation of different supervised algorithms, including MLP, to analyze tissue samples of esophago-gastric resectates [56] |
| NN | NN with 2 hidden layers used as a classifier to combine structural and spectral data for tumor detection in tongue [62] |
| AEN | Use of AEN to detect cancer in a tissue sample [69] |
| CNN | Use of CNN to detect squamous cell carcinoma between samples from different patients [67] |
| | CNN used for detection of oral cancer [61] |
| | Using specular glare in MHSI along with CNN to detect squamous cell carcinoma [68] |
| | Another study for CNN to detect squamous cell carcinoma [64] |
| | Detection of brain tumor with the aid of CNN [70] |
| | Detecting carcinoma thyroid sample with the aid of CNN [63] |
| | Different CNN models compared to one another for classifying skin cancer from patient data HIS [71] |
| GAN | GAN-inspired model used to estimate the tissue oxygen saturation [75] |
| ANN | Classifying cutaneous ulcer HIS using feedforward ANN [55] |
| ANN | ANN used to classify kidney stone tissue samples [53] |

## 4. Challenges in and Future of DL MHSI

The boom in deep learning research following the ILSVRC in 2012 has had substantial effects on a variety of disciplines [35], particularly in medical image analysis and diagnosis, as evident from the surge in papers published [78] [79]. While some domains in the medical community have already felt the impact of the applications of deep learning techniques, particularly the domain of radiology, which has severely diminished effects for the profession of practicing radiologists [80], MHSI has yet to experience a similar fate. This may simply be due to lack of broader researchers of MHSI in general. However, examining the trend observed in the number of publications for deep learning implemented in MHSI, there may be a similar outcome over the coming years. With the unprecedented advancement where general medical image analysis seemed to have reached an impasse, as deep learning methodologies augment and improvise efficiency with more accurate results [35], MHSI profession is also likely to also be impacted. Ultimately, further research for MHSI would provide the essential requirement of large available datasets, which was also an obstacle for deep learning as a whole prior to the last couple of decades.

With such deep rooted research in deep learning techniques, it is clearly evident that there is not one technique that trumps all others, as the needs and requirements vary from one situation to another. However, CNNs appear to be the prevalent choice, as the bulk of publications considered in this paper utilize it in some way or another. Meanwhile, there are still variants of architectures of CNNs implemented for a particular circumstance, such as in [37] and [68]. Although both investigate the same subject of neck and head cancer, they each use different approaches to tackle the challenge using CNNs. Issues for deep learning for MHSI also stem from several existing DL woes, one of which relates to the broader problems for classification and detection, particularly imbalance of classification for the purpose of object detection. The detection systems first perform a pixel-wide classification, as discussed in the earlier section, which typically causes the class imbalance to be biased towards non-object classes during the training process, which are easier to discern amongst the samples and may cause distortion in the overall detection process.

## 5. Conclusions

Despite the boom in deep learning, its rigorous applications in medical image analysis, and the benefit of hyperspectral imaging for medical analysis, there have not yet been any papers published that review the publications dedicated towards the implementation of deep learning for medical hyperspectral image analysis. In this paper, we discussed the deep learning techniques for medical hyperspectral imaging and relevant papers published relative to the topic within the year range of 2012-2019. In short, similar to the trend observed for deep learning since the iconic ILSVRC 2012, there was a definite boost in research and papers published for deep learning for MHSI and medical image analysis in general. We also discussed the different methodologies discussed in all the papers we found as well as what the future in DL for MHSI may look like. The trend for the majority of papers implementing DL for MHSI seems to be paved by the use of CNNs to classify the blood cells/tissue sample in order to aid cancer detection and analysis. We could only find one paper that discussed the use of FCN for segmentation rationales of retinal imaging. Lastly, there were also isolated uses of GAN and Matlab tools to determine tissue oxygenation levels that could potentially be used for detecting ischaemia (disease directly dependent on blood supply to vital organs) or even in staining of lung histology imaging.

**Conflicts of Interest:** The authors declare no potential conflict of interest and have no relevant financial motives for this article.

## Appendix A (Search Criteria)

Google Scholar was implemented to find the relative publications considered for this paper. The key search words included during our search are "deep learning" and "medical hyperspectral imaging"; we found that word searches for topics such as specific techniques yielded more inaccurate results relative to our purpose. In addition to Google Scholar, we also searched for the same search words on the publisher's own search engines for the papers considered for this article, which resulted in the discovery of papers that previously did not appear in google scholar.

## Appendix B (Datasets)

The majority of datasets utilized by the publications were obtained either directly through experimentation using various imaging systems or via contribution from hospitals, laboratories, or clinical storage. Some examples of testing systems implemented include CRI Maestro for vivo imaging system [69], more commonly used Liquid Crystal Tunable Filters (LCTF) [58] [57] [75] [59], and Hybrid endoscopic apparatus (ICL SLHSI) [65]. Several papers also used actual medical samples, such as patients undergoing surgical cancer resection [64] [37] and other contributions of local patient samples from hospitals, medical centers, and laboratories. The only largely available repository in use by a paper was BioGPS UCI repository [61].

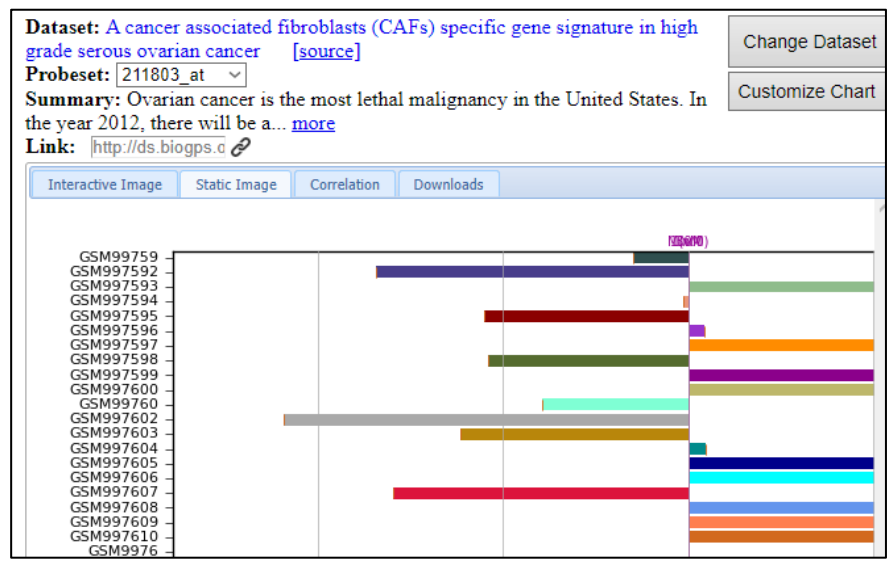

**Figure 10.** *Sample dataset from BioGPS repository: collection of genome signatures of ovarian tissue that can be utilized to identify ovarian cancer associated fibroblasts (CAFs) gene signatures [82].*

**Appendix C (Software)**

Various software packages were implemented across all the publications discussed in this paper, the majority of which implemented Tensorflow (wherever stated). Tensorflow is an open-source machine learning platform based on Python, which is supported by its vast number of libraries, tools, and community resources. The complete list of software and the papers utilizing those software can be found below-

| Software | Utilizing Publications |
|---|---|
| Statistica | [53] |
| Unscrambler (for PCA/ ANN) | [53] |
| Matlab (data processing/implementing NNs) | [58] [52] [57] [53] [54] [76] |
| Tensorflow (Python) | [58] [64] [57] [37] [65] [68] [67] [59] [63] [70] |
| Keras (Python) | [71] [59] |
| Scikit (Python) | [56] |
| Caffe | [52] |